\documentclass[natbib]{jfm}
\usepackage{bm}
\usepackage{amsmath}
\usepackage{graphicx}
\newcommand{\comment}[1]{}

\newcommand\ie{\mbox{\textit{i.e.}}}

\ifCUPmtlplainloaded \else
  \checkfont{eurm10}
  \iffontfound
    \IfFileExists{upmath.sty}
      {\typeout{^^JFound AMS Euler Roman fonts on the system,
                   using the 'upmath' package.^^J}%
       \usepackage{upmath}}
      {\typeout{^^JFound AMS Euler Roman fonts on the system, but you
                   dont seem to have the}%
       \typeout{'upmath' package installed. JFM.cls can take advantage
                 of these fonts,^^Jif you use 'upmath' package.^^J}%
      }
  \else
  \fi
\fi

\ifCUPmtlplainloaded \else
  \checkfont{msam10}
  \iffontfound
    \IfFileExists{amssymb.sty}
      {\typeout{^^JFound AMS Symbol fonts on the system, using the
                'amssymb' package.^^J}%
       \usepackage{amssymb}%
         \let\leq=\leqslant
       \let\ge=\geqslant  
      }{}
  \fi
\fi

\ifCUPmtlplainloaded \else
  \IfFileExists{amsbsy.sty}
    {\typeout{^^JFound the 'amsbsy' package on the system, using it.^^J}%
     \usepackage{amsbsy}}
    {}
\fi

\newsavebox{\astrutbox}
\sbox{\astrutbox}{\rule[-5pt]{0pt}{20pt}}

\title[Temperature spectrum generated by frictional-heating]{The temperature spectrum generated by frictional heating in isotropic turbulence}

\author[Wouter J. T. Bos]%
{Wouter\ns J.\ns  T.\ns B\ls O\ls S,\ns}

  \affiliation{LMFA-CNRS, Universit\'e de Lyon, Ecole Centrale de Lyon,
  69134 Ecully, France}

\pubyear{2013}
\volume{X}
\pagerange{1--..}
\date{2013}
\begin{document}
\maketitle

\begin{abstract}
In every turbulent flow with non-zero viscosity, heat is generated by viscous friction. This heat is then mixed by the velocity field. We consider how heat fluctuations generated this way are injected and distributed over length scales in isotropic turbulence. A triadic closure is derived and numerically integrated. It is shown how the heat fluctuation spectrum depends on the Reynolds  and Prandtl numbers.
\end{abstract}

%\pacs{47.27.Ak, 47.27.Eq, 47.27.Gs, 47.27.Jv, 47.27.Te, 47.27.Qb}

\section{Introduction}

Turbulence allows for an efficient conversion of kinetic energy into heat. In general, in the presence of external heat-sources, the influence of this heat-production is  small, and few studies have considered how these heat fluctuations are exactly redistributed over scales by the turbulent flow itself.  In a recent work, \cite{Demarinis2013}, addressed the question how this heat generation influences the wavenumber spectrum of heat fluctuations, using the Eddy-Damped Quasi-Normal Markovian (EDQNM) model, combined with an ad-hoc, spectrally local model.  Unfortunately, in the important case of a spatially uniform temperature distribution, their model predicts zero frictional heat production, which is at odds with the physics of turbulent flows, where heat fluctuations are produced as soon as the instantaneous viscous dissipation is not uniform in space. In the present manuscript we will derive an EDQNM model of the heat fluctuation production which is more generally applicable, since it is independent 
of the initial distribution of temperature fluctuations.

\section{Analysis of the problem}

The problem we will consider is conceptually the simplest possible setting in which a non-trivial effect due to the heat generation by turbulent dissipation can be observed: statistically isotropic and stationary incompressible turbulence, governed by the Navier-Stokes equations. The viscous dissipation of kinetic energy is a heat source, and the advection of this scalar, the temperature, will be considered passive, ignoring the influence of thermal expansion of the fluid elements. It will be shown later that the temperature fluctuations are in most situations small, so that this assumption seems very reasonable. The heat diffusion coefficient $\alpha$, dynamic viscosity $\mu$ and density $\rho$ are assumed to be uniform and constant. The temperature is then governed by the following equation,
\begin{equation}
 \frac{\partial \Theta}{\partial t}+u_i  \frac{\partial\Theta}{\partial x_i}=\alpha \frac{\partial^2\Theta}{\partial x_i^2}+ \frac{1}{\rho c_p}\tau_{ij}D_{ij},
\end{equation}
with $c_p$ the specific heat, which we will assume to be constant, and where the viscous stress tensor $\tau_{ij}$ is given, for a Newtonian fluid, by
\begin{equation}
 \tau_{ij}=2\mu D_{ij},
\end{equation}
and 
\begin{equation}
 D_{ij}=\frac{1}{2}\left(\frac{\partial u_i}{\partial x_j}+\frac{\partial u_j}{\partial x_i}\right).
\end{equation}
Combining these expressions we obtain (\cite{LandauBook}),
\begin{equation}
 \frac{\partial \Theta}{\partial t}+u_i  \frac{\partial\Theta}{\partial x_i}=\alpha \frac{\partial^2\Theta}{\partial x_i^2}+ \frac{\nu}{c_p}\left( \frac{\partial u_i}{\partial x_j}\frac{\partial u_i}{\partial x_j}+\frac{\partial u_i}{\partial x_j}\frac{\partial u_j}{\partial x_i} \right),
\end{equation}
with $\nu=\mu/\rho$ the kinematic viscosity. The velocity $u_i$ and temperature $\Theta$ are functions of space $\bm x$ and time $t$, but this dependence will be omitted from the notation unless more than one different space locations or time-instants appear in the same expression.  Introducing the Reynolds decomposition, $\Theta=\bar \Theta+\theta$, with $\bar \Theta$ the average temperature and $\theta$ the fluctuation, we find the following equation for the mean temperature. 
\begin{equation}\label{eq:meantemp}
 \frac{\partial \bar \Theta}{\partial t}=\frac{\nu}{c_p} \overline{\frac{\partial u_i}{\partial x_j}\frac{\partial u_i}{\partial x_j}}=\frac{\epsilon}{c_p},
\end{equation}
 with $\epsilon$ the mean dissipation rate  and where the overlined quantities denote averages. Clearly, the average temperature increases since no heat-sinks exist in our system. In a statistically stationary turbulent flow, where $\epsilon$ is constant, the mean temperature will thus increase linearly in time. 

The equation for the fluctuations is
\begin{equation}\label{eq:theta}
 \frac{\partial  \theta}{\partial t}+u_i  \frac{\partial\theta}{\partial x_i}=\alpha \frac{\partial^2\theta}{\partial x_i^2}-\frac{\nu}{c_p} \overline{\frac{\partial u_i}{\partial x_j}\frac{\partial u_i}{\partial x_j}}+\frac{\nu}{c_p}\left( \frac{\partial u_i}{\partial x_j}\frac{\partial u_i}{\partial x_j}+\frac{\partial u_i}{\partial x_j}\frac{\partial u_j}{\partial x_i}\right).
\end{equation}
The variance of the temperature fluctuations is then given by
\begin{equation}
 \frac{\partial  \overline{\theta^2}}{\partial t}=-2\alpha\overline{\frac{\partial \theta}{\partial x_i}\frac{\partial \theta}{\partial x_i}}+2\frac{\nu}{c_p}\left( \overline{\theta\frac{\partial u_i}{\partial x_j}\frac{\partial u_i}{\partial x_j}}+\overline{\theta\frac{\partial u_i}{\partial x_j}\frac{\partial u_j}{\partial x_i}}\right).
\end{equation}
Note that the second term in expression (2.11) of \cite{Demarinis2013}, is zero due to homogeneity and the last term is not complete since their basic equation (2.1) is not correct. This does not change the rest of their analysis, which did not explicitly use this information, but it is important if one wants to derive a model analytically from the governing equations, as we will do in the following section.

The temperature distribution over different lenghtscales is given by the temperature spectrum, defined such that
\begin{equation}
 \int E_\theta(k) dk =\overline{\theta^2},
\end{equation}
which implies, due to homogeneity and isotropy,
\begin{equation}
 E_\theta(k)\delta(\bm k+\bm k')=4\pi k^2\overline{\theta(\bm k)\theta(\bm k')},
\end{equation}
with $\theta(\bm k)$ the Fourier transform of the temperature fluctuation. The equation for $E_\theta(k)$ is straightforwardly derived from (\ref{eq:theta}), 
\begin{eqnarray}
\left( \frac{\partial}{\partial t} +2\alpha k^2\right) E_\theta(k)=-4\pi k^2\iint \delta(\bm k -\bm p- \bm q)ik_i \overline{u_i(\bm p)\theta(\bm q)\theta(-\bm k)}d\bm p d\bm q\nonumber\\
-4\pi k^2\frac{\nu}{c_p}\iint \delta(\bm k -\bm p- \bm q)p_iq_i \overline{u_j(\bm p)u_j(\bm q)\theta(-\bm k)}d\bm p d\bm q \nonumber\\
-4\pi k^2\frac{\nu}{c_p}\iint \delta(\bm k -\bm p- \bm q)p_iq_j \overline{u_j(\bm p)u_i(\bm q)\theta(-\bm k)}d\bm p d\bm q, \label{eq:unclosed}
\end{eqnarray}
with $u_i(\bm k)$ the Fourier transform of the velocity fluctuation. This equation is unclosed and three triple correlations need to be determined before we can integrate (\ref{eq:unclosed}). In \cite{Demarinis2013}, the first correlation, $\overline{u_i(\bm p)\theta(\bm q)\theta(-\bm k)}$ was modeled by classic EDQNM theory (\cite{Herring1982,Vignon1980}). The second correlation was modeled by the ad-hoc model,
\begin{equation}\label{wrongmodel}
-4\pi k^2\frac{\nu}{c_p}\iint \delta(\bm k -\bm p- \bm q)p_iq_i \overline{u_j(\bm p)u_j(\bm q)\theta(-\bm k)}d\bm p d\bm q\sim \frac{\nu}{c_p}k^{5/2}E(k)\sqrt{E_\theta(k)}. 
\end{equation}
As stated in the introduction, model (\ref{wrongmodel}) cannot be used in the case of a uniform temperature distribution. In that case the temperature fluctuations will stay zero forever, because the above model for the production is proportional to the spectrum $E_\theta(k)$. One could argue that the model can, perhaps, give correct results for the case in which the temperature spectrum is initially nonzero. This would however be fortuitous, since the temperature is supposed to be passive within our assumptions. It should therefore not have any influence on the injection rate of heat fluctuations. %Indeed, in the case in which an initial quantity of scalar is present in the system, one can decompose the scalar fluctuations in the system as
%\begin{equation}
% \theta(\bm x,t)= \theta^0(\bm x,t)+\tilde\theta(\bm x,t),
%\end{equation}
%in which $\theta^0(\bm x,t)$ is the initial scalar fluctuation distribution. The quantity $\tilde\theta(\bm x,t)$ is the scalar injected by Joule-dissipation. If the initial distribution $\theta^0(\bm x,0)$ is not correlated with the Joule-dissipation term,
%\begin{equation}
%2\frac{\nu}{c_p} \overline{\theta(\bm x,0)\frac{\partial u_i(\bm x',0)}{\partial x_j}\frac{\partial u_i(\bm x',0)}{\partial x_j}}=0,
%\end{equation}
%the dynamics of $\theta^0$ will remain uncorrelated with $\tilde\theta(\bm x,t)$ and the dynamics of $\tilde\theta(\bm x,t)$ can simply be described by the superposition of two uncorrelated processes. At long times the $\theta^0(\bm x,t)$ will decay to zero, whereas the contribution  $\tilde\theta(\bm x,t)$ will continue to be produced as long as a finite rate of dissipation exists in the fluid. To understand the influence of the Joule-dissipation on the scalar spectrum it is therefore enough to consider the case in which $\theta^0(\bm x,t)=0$ and the model proposed by \cite{Demarinis2013} cannot work for this case. 

We will in the following section derive an EDQNM model for the viscous production term of heat fluctuations. Subsequently we will carry out simulations to show the temperature spectrum that results from this production term. We will compare these results to those obtained by \cite{Demarinis2013}.

\section{Derivation of an EDQNM model for the  production of viscous heat fluctuations}

We use the procedure outlined in \cite{Bos2012-3} to derive the EDQNM model for the triple correlations in equation (\ref{eq:unclosed}): first we derive a Direct Interaction Approximation (DIA, \cite{KraichnanDIA}) of the triple correlation. Then we modify the time-dependence to obtain a closure expression of the EDQNM type. 
%This procedure is perhaps more formal than the standard procedure explained in a lot of investigations on EDQNM but it leads to the same final expressions. 
Note that Orszag already pointed out this relation between the Eulerian DIA equations and the EDQNM equations for isotropic turbulence in the original paper (\cite{Orszag}).  The idea behind the DIA is the following: in the triple correlations that we are interested in, the turbulent fluctuations are expanded around a state $u^{(0)}$, $\theta^{(0)}$ in which all but one particular triad are retained. The influence of this particular triad only is denoted $u^{(1)}$, $\theta^{(1)}$ and is treated as a perturbation around the state $u^{(0)}$, $\theta^{(0)}$. The expansion is then truncated at the lowest non-vanishing order. The expressions for $u^{(1)}$ and $\theta^{(1)}$ are in the present case,
\begin{eqnarray}\label{eq:u1}
u_i^{(1)}(\bm k,t)=-i\int_0^t G(k,t,s)\left(k_a\delta_{ib}+k_b\delta_{ia}-2\frac{k_ik_ak_b}{k^2} \right)u_a(\bm p,s)u_b(\bm q,s)ds
\end{eqnarray}
and
\begin{eqnarray}\label{eq:t1}
\theta^{(1)}(\bm k,t)=-i\int_0^t G_\theta(k,t,s)\left(k_a u_a(\bm p,s)\theta(\bm q,s)+k_a u_a(\bm q,s)\theta(\bm p,s)\right)ds\nonumber\\ 
-2\int_0^t G_\theta(k,t,s)\frac{\nu}{c_p}p_iq_i u_j(\bm p,s)u_j(\bm q,s) ds\nonumber\\ 
-2\int_0^t G_\theta(k,t,s)\frac{\nu}{c_p}p_iq_j u_j(\bm p,s)u_i(\bm q,s) ds
\label{eq:theta'}
\end{eqnarray}
in which $G(k,t,s)$ and $G_\theta(k,t,s)$ are the nonlinear response functions of the velocity and temperature fluctuations, respectively. Expression (\ref{eq:u1}) and the first line of (\ref{eq:t1}) are the standard DIA contributions for the velocity and the scalar (\cite{KraichnanDIA}, \cite{Roberts1961}). The second and third line of (\ref{eq:t1}) are the DIA contribution to the scalar stemming from the frictional heat source nonlinearity. 

Substituting these perturbed quantities into the expression for the triple correlations we obtain, after invoking the weak-dependence hypothesis (\cite{KraichnanDIA}),
\begin{eqnarray}\label{eq:DEtdt}
\frac{\partial E_\theta(k)}{\partial t}=T_\theta(k)+P_\theta(k)-D_\theta(k),
\end{eqnarray}
with
\begin{eqnarray}
T_\theta(k)&=&8\pi k^2\iint \delta(\bm k -\bm p- \bm q) k_ik_j \int_0^t \Phi_{ij}(\bm p,t,s) \times \nonumber\\
&&~~~~~~~~~~~~~~~~~~\left[G_\theta(k,t,s) \Phi_{\theta}(\bm q,t,s)-G_\theta(q,t,s) \Phi_{\theta}(\bm k,t,s)\right]ds d\bm p d\bm q,\nonumber\\
P_\theta(k)&=&16\pi k^2\left(\frac{\nu}{c_p}\right)^2\iint \delta(\bm k -\bm p- \bm q)  \int_0^t G_\theta(k,t,s) \times \nonumber\\
&&~~~~~~~~~~~~~~~~~~~~~~~~~~~~~\left[
(p_mq_m)^2\Phi_{ij}(\bm p,t,s)\Phi_{ij}(\bm q,t,s)\right.\nonumber\\
&&~~~~~~~~~~~~~~~~~~~~~~~~~~~~~+2p_mq_mp_iq_j\Phi_{aj}(\bm p,t,s)\Phi_{ia}(\bm q,t,s)\nonumber\\
&&~~~~~~~~~~~~~~~~~~~~~~~~~~~~~+p_i p_j q_m q_n\Phi_{mn}(\bm p,t,s)\Phi_{ij}(\bm q,t,s)
 \left.\right]
ds d\bm p d\bm q, 
\label{eq:P0}\\
D_\theta(k)&=& 2\alpha k^2 E_\theta(k),\nonumber
\end{eqnarray}
where the letters $T,P,D$ denote Transfer, Production and Diffusion, respectively. In this expression we have introduced the quantities
\begin{eqnarray}
\Phi_{ij}(\bm k,t,s)\delta(\bm k+\bm k')=\overline{u_i(\bm k,t)u_j(\bm k',s)},~~~~~~~
\Phi_{\theta}(\bm k,t,s)\delta(\bm k+\bm k')=\overline{\theta(\bm k,t)\theta(\bm k',s)}. 
\end{eqnarray}
We consider the isotropic, mirror-symmetric case so that, introducing the time correlation functions $R(k,t,s)$ and $R_\theta(k,t,s)$,
\begin{eqnarray}
\Phi_{ij}(\bm k,t,s)=\Phi_{ij}(\bm k,t,t)R(k,t,s),~~~~~~~
\Phi_{\theta}(\bm k,t,s)=\Phi_{\theta}(\bm k,t,t)R_{\theta}(k,t,s), 
\end{eqnarray}
we can write
\begin{eqnarray}
\Phi_{ij}(\bm k,t,s)=\frac{E(k)}{4\pi k^2}\left(\delta_{ij}-\frac{k_ik_j}{k^2}\right) R(k,t,s),~~~~~~~
\Phi_{\theta}(\bm k,t,s)=\frac{E_\theta(k)}{4\pi k^2} R_\theta(k,t,s). 
\end{eqnarray}
The expressions for $T_\theta(k)$ and $P_\theta(k)$ can be rewritten, after some algebraic manipulations, as
\begin{eqnarray}
T_\theta(k)&=&\iint_\Delta \Theta^I(k,p,q)(1-z^2)k^3q^2E(p)\left[\frac{E_\theta(q)}{q^2}-\frac{E_\theta(k)}{k^2}\right]\frac{dpdq}{pq},\nonumber\\
P_\theta(k)&=&2\left(\frac{\nu}{c_p}\right)^2\iint_\Delta \Theta^{II}(k,p,q)(4x^4-3x^2+1)kp^2q^2E(p)E(q)\frac{dpdq}{pq} \label{eq:PEDQNM}
\end{eqnarray}
in which
\begin{eqnarray}
\Theta^I(k,p,q)=\int_0^t G_\theta(k,t,s)R_\theta(q,t,s)R(p,t,s)ds\nonumber\\
\Theta^{II}(k,p,q)\int_0^t G_\theta(k,t,s)R(q,t,s)R(p,t,s)ds.
\end{eqnarray}
and in which $\Delta$ indicates the domain in the $pq$-plane in which
$k,p,q$ can form a triangle (in other words $|p-q|\leq k \leq|p+q|$) and $x,y,z$ are given by
\begin{eqnarray}
x=-p_iq_i/(pq)\nonumber\\
y=k_iq_i/(kq)\nonumber\\
z=k_ip_i/(kp).\label{eq:xyz}
\end{eqnarray}
Assuming a fluctuation-dissipation relation for the two-time quantities and exponentially decaying time-correlations, 
\begin{eqnarray}
R(k,t,s)=G(k,t,s)=\exp(-\eta(k)(t-s)) ~~\textrm{for $t\ge s$},\\
R_\theta(k,t,s)=G_\theta(k,t,s)=\exp(-\eta_\theta(k)(t-s)) ~~\textrm{for $t\ge s$},
\end{eqnarray}
these expressions simplify, in the long-time limit, to 
\begin{eqnarray}
\Theta^I(k,p,q)=\frac{1}{\eta_\theta(k)+\eta_\theta(q)+\eta(p)}\nonumber \\
\Theta^{II}(k,p,q)=\frac{1}{\eta_\theta(k)+\eta(p)+\eta(q)}.
\end{eqnarray}
It is at this point that the choice of the form of the typical frequencies $\eta$ and $\eta_\theta$ will determine the inertial range scaling of the spectra. We will choose the standard forms (\cite{Herring1982}),
\begin{eqnarray}
 \eta(k)=1.3\sqrt{\int_0^k s^2 E(s)ds}+\nu k^2 \textrm{~~~and~~~} \eta_\theta(k)=\alpha k^2.
\end{eqnarray}
We note here that the expression for $T_\theta(k)$ is the standard EDQNM closure for scalar fluctuations (\cite{Herring1982,Vignon1980}). The expression for the viscous heat production term $P_\theta(k)$ is new. In principle, the presence of the frictional heating term could change the behaviour of the scalar frequency $\eta_\theta(k)$. This possibility is not explored in the present investigation.

The equation for the energy spectrum is equivalently closed by the EDQNM model (\cite{Orszag}),
\begin{eqnarray}\label{eq:dEdt}
\left( \frac{\partial}{\partial t} +2\nu k^2\right) E(k)&=&\iint_\Delta \Theta(k,p,q)(xy+z^3)k^3p^2E(q)\left[\frac{E(p)}{p^2}-\frac{E(k)}{k^2}\right]\frac{dpdq}{pq}+F(k),\nonumber\\
\end{eqnarray}
with $F(k)$ a constant forcing term acting in the small wavenumber range and 
\begin{eqnarray}
\Theta(k,p,q)=\frac{1}{\eta'(k)+\eta'(p)+\eta'(q)}.
\end{eqnarray}
with 
\begin{eqnarray}
\eta'(k)=0.36\sqrt{\int_0^k s^2 E(s)ds}+\nu k^2.
\end{eqnarray}
It was shown in the past that the variance of the dissipation-rate fluctuations, involving fourth order correlations of the velocity field, was not correctly captured by DIA related approaches (\cite{Chen1989}). The present model involves the correlation between the dissipation-rate fluctuation and the temperature fluctuation. This is a mixed velocity-temperature third order correlation. In general, it seems that DIA more reliably represents third than fourth order correlations, but a validation of the model (\ref{eq:PEDQNM}) by direct numerical simulation would be valuable, in particular since the modeled correlation involves the fluctuation of the dissipation rate.

\section{Heat production at unit Prandtl number}

\subsection{Preliminary analysis of the closure expression}

The goal is here to first derive expressions for the heat-fluctuation spectrum generated by frictional heating, by considering the evolution equation (\ref{eq:DEtdt}) and the EDQNM closure for the frictional heating term $P_\theta(k)$ in this equation. Then we will verify these analytical results by numerically integrating the closure equations. For our analytical considerations we will assume the energy spectrum to be constant in time and to obey Kolmogorov scaling, $E(k)\sim \epsilon^{2/3}k^{-5/3}$, from the wavenumber corresponding to the integral scale down to the Kolmogorov wavenumber $k_\eta\sim\epsilon^{1/4}\nu^{-3/4}$. 

%Intuitively, we understand that the frictional heating is generated mainly at large wavenumbers, $k\approx k_\eta$, where the gradients become important enough to induce a significant friction. 
 To evaluate the convolution integral in (\ref{eq:P0}), there are three distinct cases to consider. Firstly, the convolution integral is dominated by local interactions. Secondly, the integral is dominated by infrared nonlocal interactions $p\ll k\approx q$. Thirdly the integral is dominated by ultraviolet nonlinear interactions $k\ll p\approx q$. We have verified the convolution integral analytically for all three types of contributions and we have found that in the case of a very long inertial range, the nonlocal interactions contribute insignificantly compared to the local interactions. We will therefore only consider the approximation of (\ref{eq:DEtdt}) for the case of local interactions. In that case (\ref{eq:P0}) can be approximated by its dimensional estimate,
\begin{equation}
 P_\theta(k)\sim \left(\frac{\nu}{c_p}\right)^2 k^5 E(k)^2 \tau(k),
\end{equation}
where $\tau(k)$ is the local estimate of the triad interaction time $\Theta^{II}(k,p,q)$. If we suppose 
\begin{equation}\label{eq:tauk}
\tau(k)\sim \epsilon^{-1/3}k^{-2/3}, 
\end{equation}
and we assume Kolmogorov-scaling we find 
\begin{equation}\label{eqPposs1}
 P_\theta(k)\sim \left(\frac{\nu}{c_p}\right)^2 \epsilon k.
\end{equation}
In the case of unit and large Prandtl number, the diffusion will be small in the inertial-convective range, and we expect a balance to establish between the production term and the transfer term,
\begin{equation}\label{eq:PisT}
 P_\theta(k)\approx T_\theta(k).
\end{equation}
This balance will indeed be observed in the numerical evaluation of the closure. In this range, it was shown (\cite{Rubinstein2013}) that the scalar transfer can be qualitatively predicted by a Kovaznay-type of gradient model,
\begin{eqnarray}\label{eq:Kovaznay}
T_\theta(k)\sim \frac{\partial }{\partial k} \left(E_\theta(k)E(k)^{1/2}k^{5/2} \right), 
\end{eqnarray}
we find from (\ref{eqPposs1}), (\ref{eq:PisT}) and (\ref{eq:Kovaznay}),
\begin{eqnarray}\label{eq:EthighP}
E_\theta(k)\sim \left(\frac{\nu}{c_p}\right)^2 \epsilon^{2/3}k^{1/3}.
\end{eqnarray}
Integrating this spectrum up to the kolmogorov scale we have,
\begin{eqnarray}\label{eq:t2Re}
\overline{\theta^2}\sim \frac{\epsilon \nu}{c_p^2}\sim R_\lambda^{-2}.
\end{eqnarray}
In a steady state, the total production will be balanced by the scalar dissipation, so that,
\begin{equation}
\int P_\theta(k)dk=\epsilon_\theta \equiv\int D_\theta(k)dk.
\end{equation}
Assuming a scaling of the production spectrum of the form (\ref{eqPposs1}) up to the Kolmogorov scale $k_\eta$, we have
\begin{equation}\label{eq:epsTlocal}
\epsilon_\theta\sim \frac{\epsilon^{3/2}\nu^{1/2}}{c_p^2},
\end{equation}
which shows that in the limit $R_\lambda \rightarrow \infty$,
\begin{equation}
\epsilon_\theta\sim R_\lambda^{-1}.
\end{equation}
This means that in the limit of vanishingly small viscosity, both the mean-square heat fluctuations and their dissipation tend to zero.

\subsection{Numerical integration of the closure for unit Prandtl number\label{sec:resultsUnity}}

\begin{figure}
\begin{center}
\includegraphics[width=0.5\linewidth,angle=0]{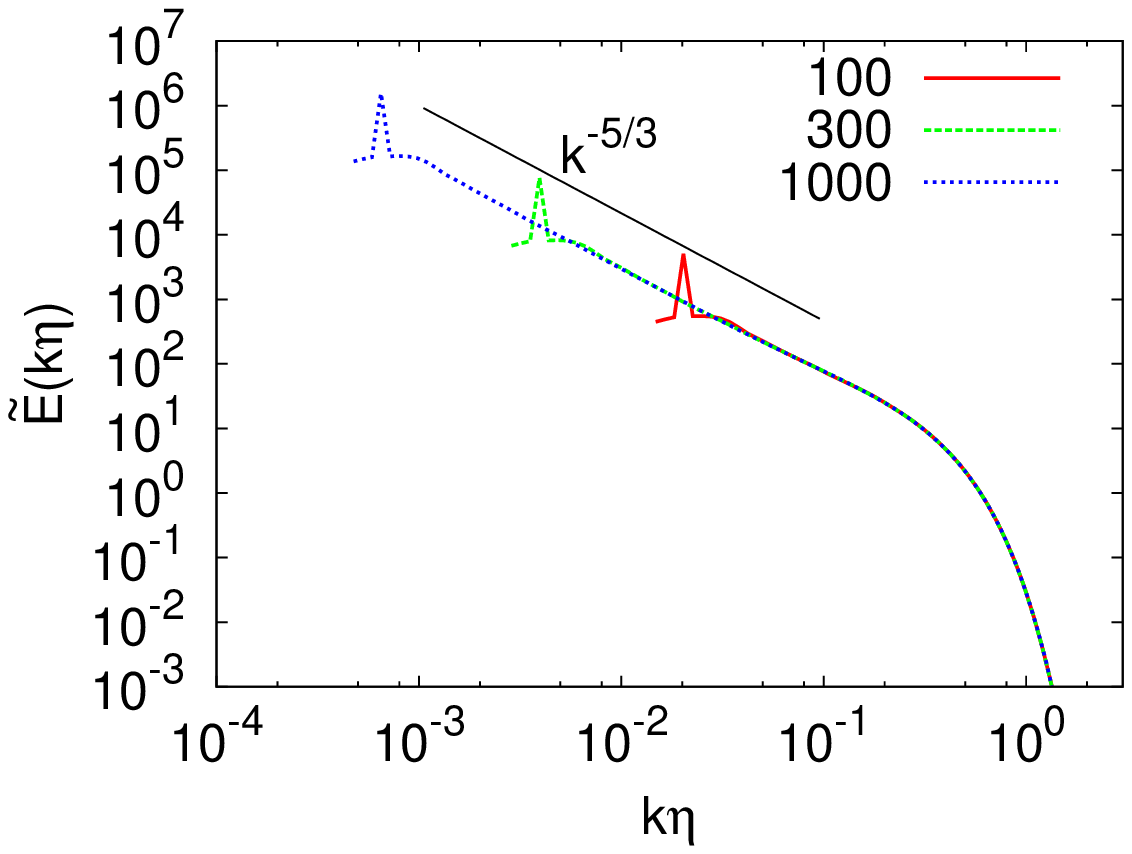}~
\includegraphics[width=0.5\linewidth,angle=0]{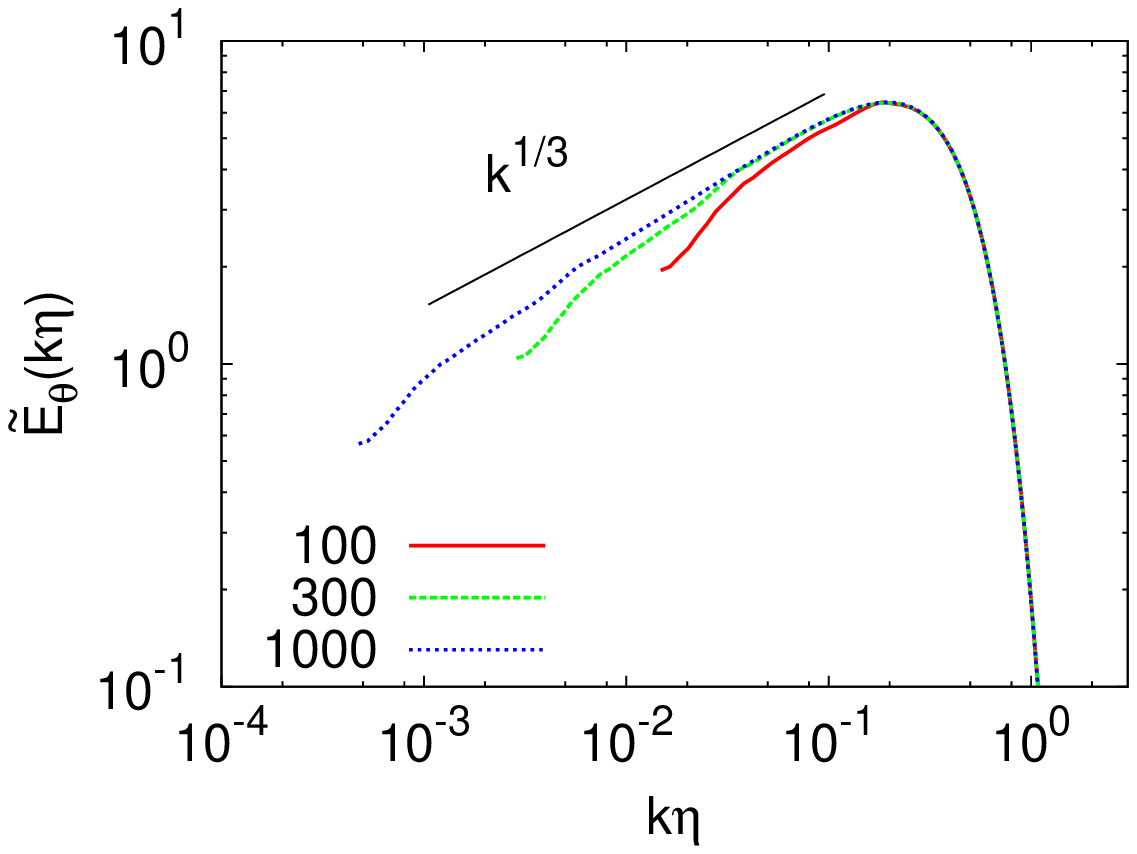}
\caption{Left: Energy spectrum, normalized by Kolmogorov variables, for three different Reynolds numbers. Right: corresponding temperature fluctuation spectrum, generated by frictional heating ($Pr=1$). }
\label{Fig:1}
\end{center}
\end{figure}

\begin{figure}
\includegraphics[width=0.5\linewidth,angle=0]{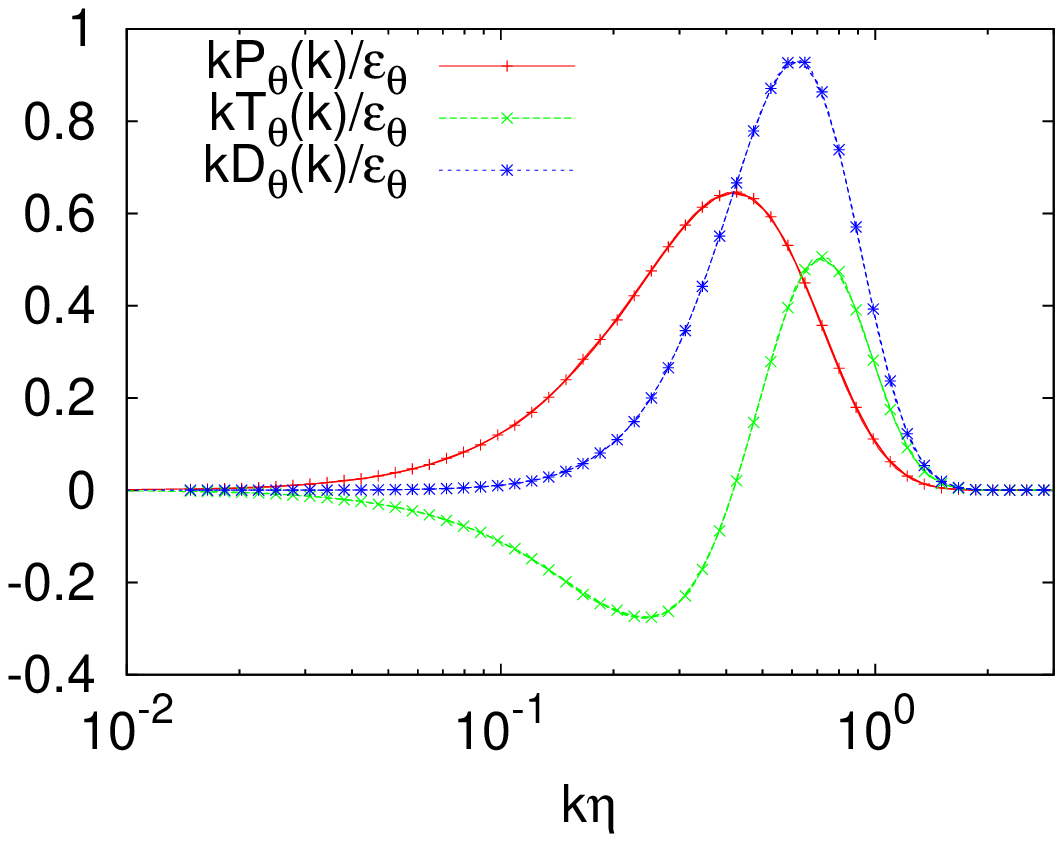}~
\includegraphics[width=0.5\linewidth,angle=0]{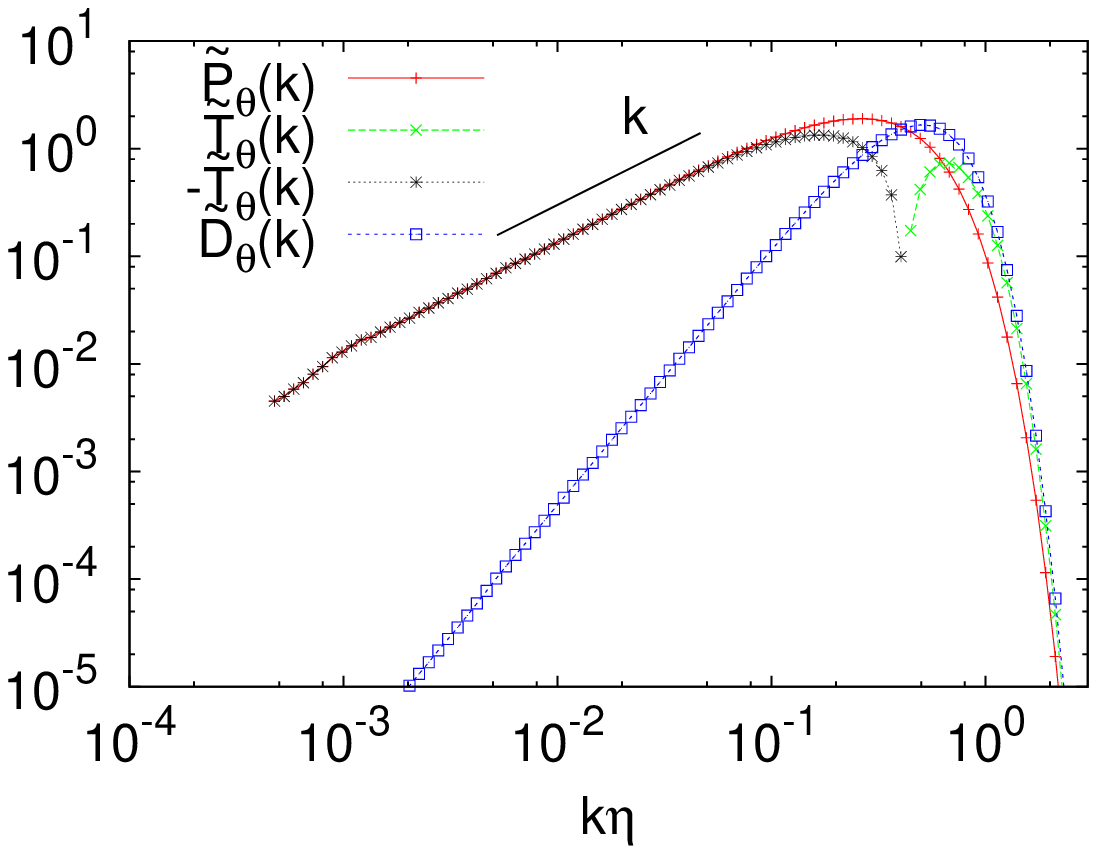}
\caption{Left: different terms in the evolution equation of the temperature spectrum  ($Pr=1$): transfer $T_\theta(k)$, production by frictional dissipation $P_\theta(k)$ and  dissipation of temperature fluctuations $D_\theta(k)$. Right: logarithmic representation. }
\label{Fig:2}
\end{figure}

\begin{figure}
\includegraphics[width=0.5\linewidth,angle=0]{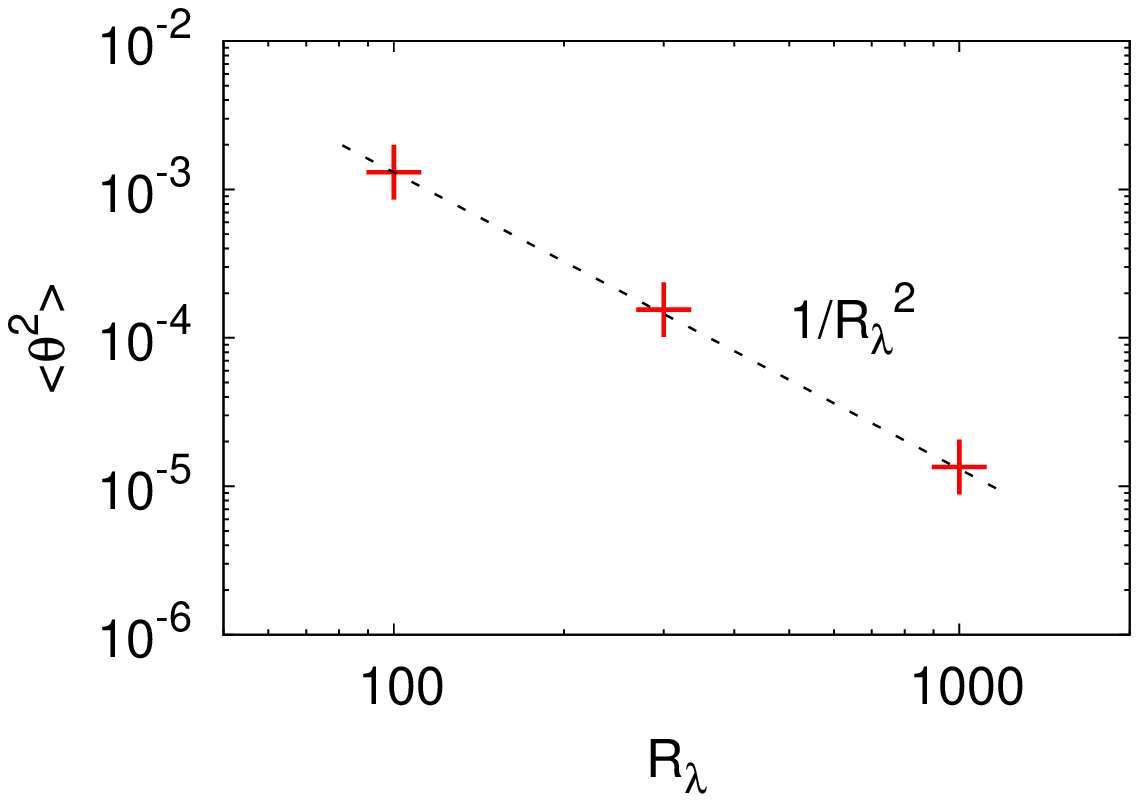}
\includegraphics[width=0.5\linewidth,angle=0]{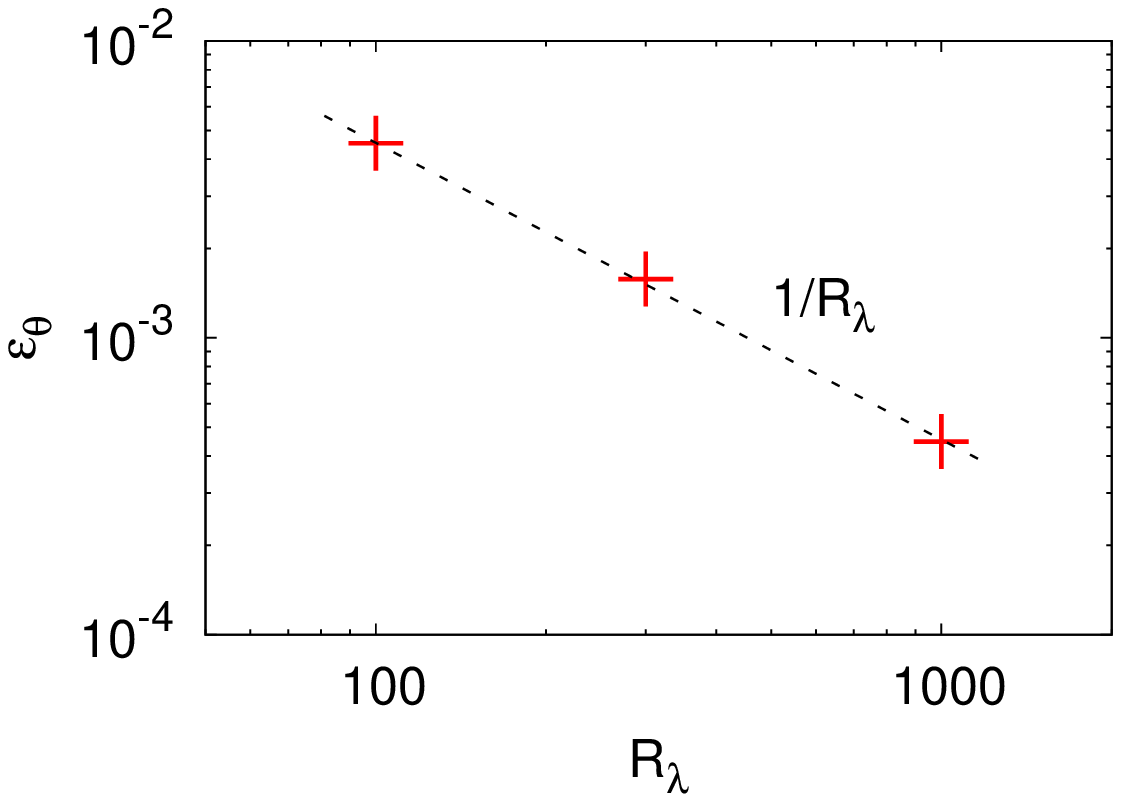}
\caption{ Reynolds number dependence of the mean-square heat fluctuations (left), and of the dissipation of heat-fluctuations (right) at $Pr=1$.}
\label{Fig:epsT}
\end{figure}

Equations (\ref{eq:DEtdt}) and (\ref{eq:dEdt}) are integrated numerically on a logarithmically spaced grid with 22 wavenumbers per decade for Reynolds number $R_\lambda=100,~300,~1000$. The specific heat $c_p$ is kept constant, since its variation only induces a trivial change on the results. Indeed, since the advection equation is linear and depends linearly on $c_p^{-1}$, the temperature spectrum will simply be proportional to $c_p^{-2}$. The initial temperature spectrum is zero. All wavenumber spectra in the remainder of this results section are normalized using Kolmogorov-Corrsin-Obukhov variables, \ie, using the variables $\nu$, $\epsilon$ and $\epsilon_\theta$. The normalized spectra are indicated by a tilde. For instance, the normalized energy spectrum is given by
\begin{equation}
 \tilde E(k)=\nu^{-5/4}\epsilon^{-1/4}E(k),
\end{equation}
and $k$ is normalized by the Kolmogorov lengthscale $\eta=\nu^{3/4}\epsilon^{-1/4}$. 

Since the velocity field contains a large scale forcing, the energy spectrum will develop towards a steady state. This long-time asymptotic energy spectrum is shown in Figure \ref{Fig:1}, left. The energy spectra, normalized using Kolmogorov variables, collapse perfectly at high wavenumbers. The temperature spectrum which develops asymptotically due to the viscous heat production is shown in  Figure \ref{Fig:1}, right. It is an increasing function of the wavenumber, roughly proportional to $k^{1/3}$ and the spectra rapidly fall off in the far dissipation range. For different Reynolds numbers the temperature spectra collapse, using Kolmogorov-Corrsin-Obukhov variables,
\begin{eqnarray}
 \tilde E_\theta(k)=(\epsilon_\theta\nu^{5/4}\epsilon^{3/4})^{-1}E_\theta(k).
\end{eqnarray}
In Figure \ref{Fig:2} we show the three different contribution terms in the evolution equation of the temperature spectrum: transfer $T_\theta(k)$, production by frictional heating $P_\theta(k)$ and  dissipation of temperature fluctuations $D_\theta(k)$.  We observe that the main part of the production is locally (in wavenumber) dissipated by the diffusion. The transfer term allows for a small transfer of temperature variance towards smaller scales. But since the production takes mainly place in the diffusive range (at unit Prandtl number), the fluctuations do not survive long enough to be strained significantly by the velocity field, and the transfer is therefore confined to a small wavenumber span in the dissipation range. This picture will change at large Prandtl number. In Figure \ref{Fig:2}, right, we show the same contributions, but in a logarithmic representation. This allows to more closely focus on the 
small and intermediate wavenumber ranges.  In the inertial range, the production term is proportional to $k$.
The production is largely dominant over the diffusion $D_\theta(k)$ in this range and it is therefore the transfer term which balances the production, as assumed in expression (\ref{eq:PisT}). The wavenumber dependence of $P_\theta(k)$ and $E_\theta(k)$ are as predicted in expressions (\ref{eqPposs1}) and (\ref{eq:EthighP}), which were obtained by assuming that the convolution integral in (\ref{eq:P0}) is dominated by local interactions. A further consequence of this assumption is that the mean-square temperature fluctuations should be proportional to $R_\lambda^{-2}$. The dissipation of temperature fluctuations should be inversely proportional to the Reynolds number. This is indeed observed in Figure \ref{Fig:epsT}.

\section{Temperature spectrum in the inertial-diffusive and viscous-convective range \label{sec:inertial}}

\subsection{Phenomenology for small and large Prandtl numbers}

The energy spectrum rapidly decays for $k>k_\eta$. Since the convolution integral in (\ref{eq:PEDQNM}) contains interactions with $E(p)$ and $E(q)$ under the constraint that  $|p-q|\leq k \leq|p+q|$, negligible production will be observed beyond $k=2k_\eta$. For very high Prandtl number we therefore expect a Batchelor-type phenomenology  in which the scalar fluctuations are strained by the smallest available velocity scale, around $k_\eta$ (\cite{Batchelor1959}). This scaling is given by
\begin{equation}
 E_\theta(k)\sim \epsilon_\theta \sqrt{\frac{\nu}{\epsilon}}k^{-1}.
\end{equation}
Substituting $\epsilon_\theta$ by expression (\ref{eq:epsTlocal}), we have
\begin{equation}\label{eq:EtBatch}
 E_\theta(k)\sim \frac{\epsilon\nu}{c_p^2}k^{-1}.
\end{equation}
For the opposite case, in which the Prandtl number is very small, the production term will be balanced by the diffusion term. The production term, was approximated by 
\begin{equation}
 P_\theta(k)\sim \left(\frac{\nu}{c_p}\right)^2 k^5 E(k)^2 \tau(k),
\end{equation}
assuming dominance of the local interactions in the generation of the heat fluctuations. However, the triad-correlation timescale is in this case not given by (\ref{eq:tauk}), since at these scales the dynamics are now dominated by the diffusion of heat, so that
\begin{equation}
\tau(k)\sim 1/(\alpha k^2),
\end{equation}
and the production spectrum in the inertial-diffusive range will then be given by
\begin{equation}
 P_\theta(k)\sim \left(\frac{\nu}{c_p}\right)^2 \frac{\epsilon^{4/3}}{\alpha} k^{-1/3}.
\end{equation}
Since the diffusion spectrum is
\begin{equation}
 D_\theta(k)=2\alpha k^2 E_\theta(k),
\end{equation}
we find from 
\begin{equation}\label{eq:lowP}
 P_\theta(k)\approx D_\theta(k),
\end{equation}
that
\begin{equation}\label{eq:EtlowP}
 E_\theta(k)\sim \left(\frac{\nu}{c_p}\right)^2 \frac{\epsilon^{4/3}}{\alpha^2} k^{-7/3}.
\end{equation}

\subsection{Numerical integration of the closure for small and large Prandtl number\label{sec:resultsPr}}

\begin{figure}
\includegraphics[width=0.5\linewidth,angle=0]{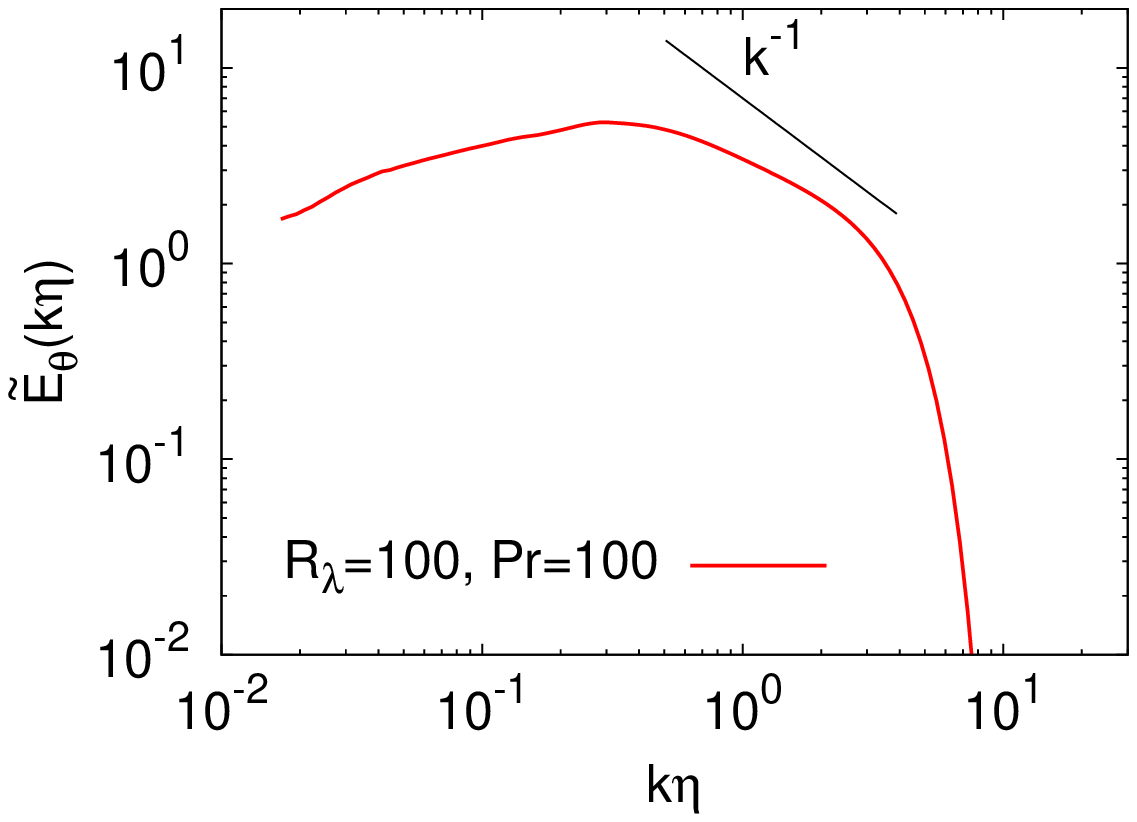}~
\includegraphics[width=0.5\linewidth,angle=0]{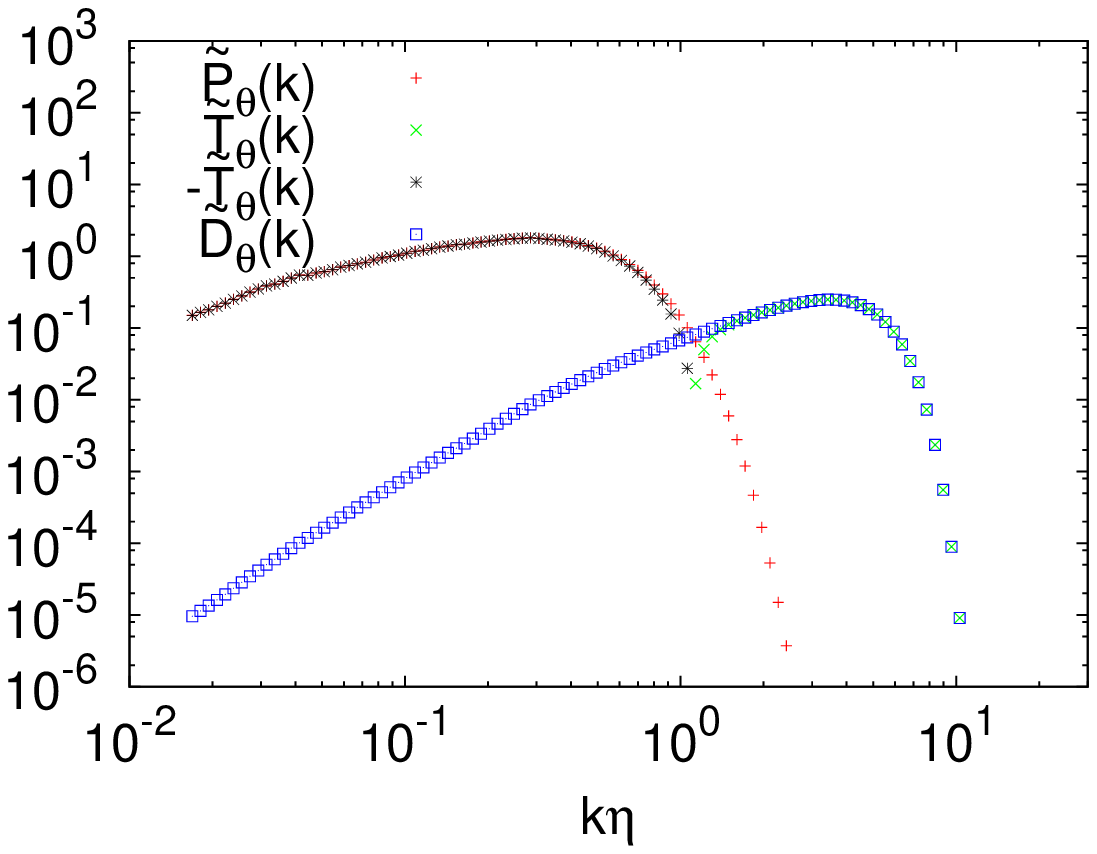}
\caption{Left: $E_\theta(k)$ for $R_\lambda=100$, $Pr=100$. Right, transfer $T_\theta(k)$, production $P_\theta(k)$ and  dissipation spectrum $D_\theta(k)$. }
\label{Fig:4}
\end{figure}

The results for a large Prandtl number simulation ($Pr=100$, $R_\lambda=100$) are shown in Figure \ref{Fig:4}. The $k^{-1}$ spectrum starts to appear but is not yet fully developed at this Prandtl number. Since the peak of the production mechanism remains fixed around $k_\eta$, independently of the Prandtl number, there is no reason why it should not become more pronounced at higher Prandtl number. The balance of diffusion and transfer is observed in Figure \ref{Fig:4}, right, for wavenumbers larger than $k_\eta$ where the energy spectrum drops rapidly.  

\begin{figure}
\includegraphics[width=0.5\linewidth,angle=0]{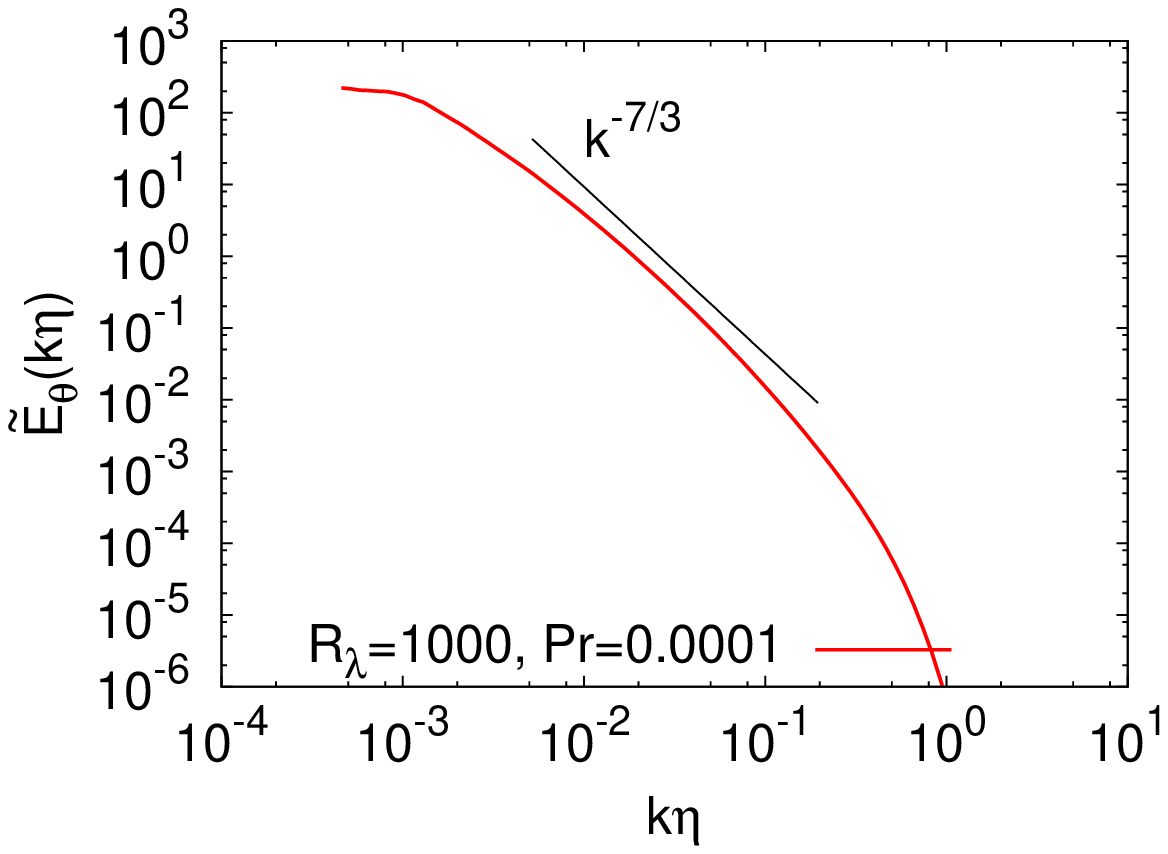}~
\includegraphics[width=0.5\linewidth,angle=0]{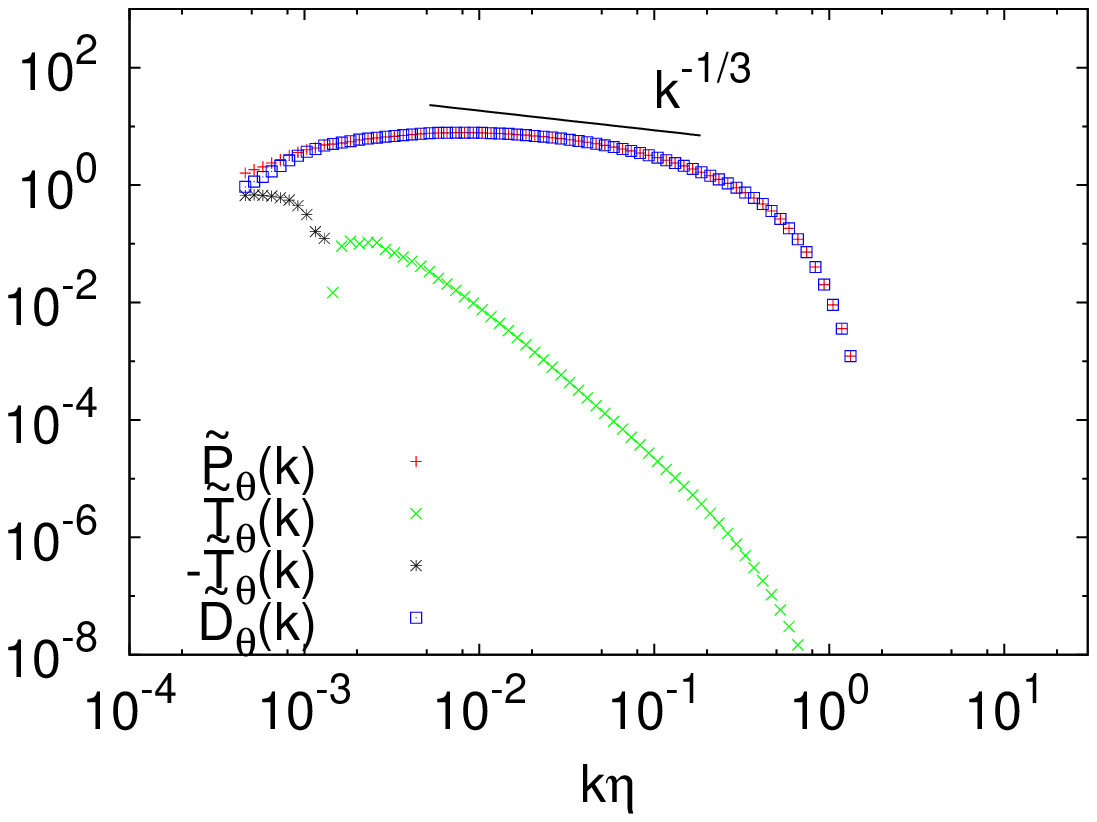}
\caption{Left: $E_\theta(k)$ for $R_\lambda=1000$, $Pr=10^{-4}$. Right, transfer $T_\theta(k)$, production $P_\theta(k)$ and  dissipation spectrum $D_\theta(k)$. }
\label{Fig:3}
\end{figure}

Figure \ref{Fig:3}, left, shows that, at low Prandtl number, the temperature spectrum is indeed approximately proportional to $k^{-7/3}$, in agreement with expression (\ref{eq:EtlowP}).   The balance between the diffusion term and the heat fluctuation source term holds, as is observed in Figure \ref{Fig:3}, right.  The production spectrum is not exactly given by a powerlaw proportional to $k^{-1/3}$, which is consistent with the fact that the temperature spectrum is not described by a pure powerlaw either.

\section{Comparison with previous model predictions}

It was shown in the previous sections that according to the closure we derived, the generation of heat fluctuations is dominated by local interactions in wavenumber space. This allowed us to show that a simple, phenomenological model of the production is 
\begin{eqnarray}\label{eq:recallmodel}
 P_\theta(k)\sim \left(\frac{\nu}{c_p}\right)^2 k^5 E(k)^2 \tau(k),
\end{eqnarray}
with 
\begin{equation}
\tau(k)\sim\left\{\begin{array}{ll}
\epsilon^{-1/3}k^{-2/3}& \textrm{ for } Pr\ge 1, k\eta\ll 1\\
1/(\alpha k^2)&  \textrm{ for }  Pr\ll 1, k\eta\ll 1\\
\end{array}\right..
\end{equation}
Comparing the form of the above model to the model of \cite{Demarinis2013}, the main difference is the absence of the heat fluctuation spectrum in (\ref{eq:recallmodel}). In the case of the model of \cite{Demarinis2013}, the production term cannot be determined independently from the temperature fluctuation distribution. All cases we considered in the present work would thus yield zero temperature fluctuations using that model. However, in the particular case of an initial heat-fluctuation spectrum, in the absence of further heat-sources and for very long times, it is possible to compare the predictions of the model of \cite{Demarinis2013} with the present case. For instance in the inertial-convective range we expect a balance between nonlinear-transfer of scalar fluctuations and production. Combining expression (\ref{eq:PisT}), (\ref{eq:Kovaznay}) and (\ref{wrongmodel}), we find in this particular case for the temperature fluctuation spectrum in the inertial-convective range (assuming Kolmogorov scaling),
\begin{equation}
 E_\theta(k)\sim \left(\frac{\nu}{c_p}\right)^2 \epsilon^{2/3}k^{1/3}. 
\end{equation}
If we consider the low Prandtl number case, balance (\ref{eq:lowP}) should hold in the inertial range and the temperature spectrum will be proportional to 
\begin{equation}
 E_\theta(k)\sim \left(\frac{\nu}{c_p}\right)^2 \alpha^{-2}\epsilon^{4/3}k^{-7/3},
\end{equation}
as also proposed by \cite{Demarinis2013}. The predictions of their model for these particular cases are therefore in agreement with the results of our EDQNM model. This might seem surprising, given the fact that the model does not give physical results for the basic case in which no initial fluctuations are present. We note however that all the models of the form 
\begin{equation}\label{eq:modgamma}
P_\theta(k)\sim E_\theta(k)^\gamma \left(\frac{\nu}{c_p}\right)^{2(1-\gamma)}E(k)^{3/2-\gamma} k^{7/2-2\gamma} 
\end{equation}
will give the same steady-state inertial-convective range behavior, independent of the value of $\gamma$. The only choice predicting a non-zero production in the absence of initial fluctuations is $\gamma=0$. The value that will give correct scaling in the inertial-diffusive range, at low Prandtl number is $\gamma=1/2$ corresponding to the model of \cite{Demarinis2013}. It is therefore not possible to find a model of the form (\ref{eq:modgamma}) which gives behaviour in agreement with both (\ref{eq:EthighP}) and (\ref{eq:EtlowP}) and which predicts nonzero production in the absence of initial fluctuations. The simplest phenomenological model that satisfies these criteria is (\ref{eq:recallmodel}), involving a timescale which can mimic both diffusive and inertial behaviour.

We note that as soon as initial heat fluctuations are present in the considered flow, as is a prerequisite for (\ref{wrongmodel}) to generate heat, the distribution of these fluctuations will determine the instantaneous heat generation and all transients will be affected by this. The influence of these unphysical transients can be very long if the initial heat fluctuations are present in the largest flow scales, which decay slowly.

\section{Conclusion}

An EDQNM model for the heat-production term due to viscous friction is derived. Temperature wavenumber spectra are deduced from this closure.  In particular, it is shown that the heat-production is mainly generated through local triad interactions. The heat production spectrum is then proportional to $k$. Balancing this production spectrum against nonlinear transfer, we find the heat-spectrum for unit and large Prandtl number in the inertial-convective subrange. In the very diffusive case (very small Prandtl number), the production is in equilibrium with the diffusivity spectrum, leading to a different wavenumber dependence.  The predictions are
\begin{equation}\label{Expressions2-b}
E_\theta(k)\sim\left\{\begin{array}{ll}
\left(\frac{\nu}{c_p}\right)^2 \epsilon^{2/3}k^{1/3}&  \textrm{ for }  Pr\ge 1, k\eta\ll 1\\
\left(\frac{\nu}{c_p}\right)^2 \frac{\epsilon^{4/3}}{\alpha^2} k^{-7/3}& \textrm{ for } Pr\ll 1, k\eta\ll 1\\
\frac{\epsilon\nu}{c_p^2}k^{-1}& \textrm{ for } Pr\gg 1, k\eta\gg 1.\\
\end{array}\right.
\end{equation}

Now that we have obtained these expressions for the heat-fluctuation spectra, it is possible to estimate how large the fluctuations are in reality. Expression  (\ref{eq:t2Re}) shows that, assuming Kolmogorov scaling, we have
\begin{equation}
\epsilon\sim \frac{\overline{\theta^2}c_p^2}{\nu}.
\end{equation}
Let us consider typical values for $c_p$  and $\nu$. For a gas such as air, $c_p\approx 10^3 J kg^{-1}K^{-1}$, $\nu\approx 10^{-5}m^2s^{-1}$. In order to observe heat-fluctuations of the order of $10^{-3}K$, a dissipation rate of the order of $10^5~ W kg^{-1}$ is needed. Combining this with equation (\ref{eq:meantemp}), this implies that the mean temperature of the gas increases at a rate of $10^2 Ks^{-1}$. For a typical liquid such as water, this increase is even larger. This example illustrates  the smallness of the temperature variance in practice. It does not seem to us that in a practical situation in which the temperature of a fluid increases by $100 K$ every second, fluctuations of the order of a millikelvin are of major importance.

Note that the scalings (\ref{Expressions2-b}) are observed approximately from the numerically integrated closure expressions. We stress that these are the results which are given within the framework of the EDQNM assumptions and the model cures for the defect of a previous model proposition which excluded heat-fluctuation generation in the absence of initial fluctuations (\cite{Demarinis2013}). A validation of the EDQNM model for the heat production and a verification of the proposed scalings (\ref{Expressions2-b}) using high resolution direct numerical simulations of isotropic turbulence would certainly be valuable,  in particular since DIA and EDQNM might have difficulties in representing correlations involving fluctuations of the dissipation rate.

\section*{Ackowledgments}

The author thanks Robert Rubinstein for his invaluable comments and acknowledges discussions with Marcello Meldi, Sergio Chibbaro and Pierre Sagaut. 

%\bibliographystyle{myJFM}
%\bibliography{/home/bos/PUBLI/biblio}

\end{document}